\pdfoutput=1
\documentclass[5p]{elsarticle}

\usepackage{lineno,hyperref}
\modulolinenumbers[5]

\journal{Journal of \LaTeX\ Templates}

\usepackage{amsmath}
\newcommand{\CP}{$C\!P$}
\usepackage[usenames, dvipsnames]{xcolor}
\newcommand{\SLV}[1]{\boldsymbol{#1}}
\usepackage{lipsum}
\usepackage{braket} 
\usepackage[capitalise]{cleveref}
\usepackage{booktabs}

\begin{document}

\begin{frontmatter}

\title{New-physics searches in \CP{}-violating $\eta$ muonic decays}

\author{Pablo Sanchez Puertas\fnref{myfootnote}}
\address{Institut de F{\'i}sica d'Altes Energies (IFAE),\\ The Barcelona Institute of Science and Technology, \\ Universitat Aut{\'o}noma de Barcelona, E-08193 Bellaterra (Barcelona), Spain}
\fntext[myfootnote]{psanchez@ifae.es}

%
%

\begin{abstract}
In this work we investigate the possibility to observe \CP{}-violation in $\eta$ decays containing muons at the proposed REDTOP experiment. 
Employing the SMEFT to parametrize the new-physics \CP{}-violating effects, we find that a single operator exists for which is possible to observe \CP{}-violation at REDTOP in $\eta\to\mu^+\mu^-$ decays, while evading bounds from the neutron electric dipole moment and $D^-\to\mu\bar{\nu}$ meson decays.

\end{abstract}


\end{frontmatter}


\section{Introduction}

The single origin of \CP{} violation in the Standard Model (SM) of particle physics is rooted in the CKM matrix and connected to the Jarlskog invariant~\cite{Jarlskog:1985cw}---a source which is known to be small and to appear in electroweak decays alone. 
As such, the observation of \CP{} violation in processes dominated by the strong or electromagnetic interactions would be an almost unambiguous proof of new physics.\footnote{This contrasts with $K$ mesons that decay weakly, implying that \CP{}-violating new physics effects could be of the same order as the SM. Indeed, there is an actual controversy on whether new physics effects have been found in this sector or not~\cite{Buras:2018wmb,Gisbert:2017vvj}.} 
Such a possibility is brought by $\eta$ meson decays: these proceed mainly via the strong and electromagnetic interactions and correspond to well-defined quantum numbers with $I^G(J^{PC}) = 0^+(0^{-+})$. Of course, this comes at a price to pay: the necessity of a large sample of such decays. For this reason, an experimental Collaboration named REDTOP has been proposed~\cite{Gatto:2016rae}. The Collaboration expects to produce around $10^{12}$ $\eta$ mesons and its main target is to measure \CP{} violation in $\eta$ decays. To that end, part of the program focus on the ability of the experiment to measure the polarization of the muons in $\eta$ decays, that carries important information about \CP{} violation.

However, it is very important for the preliminary studies to assess whether the proposed experiment is competitive with different observables that set stringent bounds, as it is the case here when connecting with electric dipole moments (EDMs). In this work~\cite{Sanchez-Puertas:2018tnp}, we carry on such a study, for which we will assumme that new physics is heavy enough to use the SMEFT. The basic ingredients we need and the hadronization details are outlined in \cref{sec:semeft}, while the muon polarization in different $\eta$ muonic decays is computed in \cref{sec:pol}---where the experimental sensitivity is assessed. Finally, in \cref{sec:bounds}, we study the bounds that different observables put on these processes (including a new processes not considered originally in \cite{Sanchez-Puertas:2018tnp}), finding that is still possible to observe \CP{}-violation in $\eta\to\mu^+\mu^-$ decays.

\section{The SMEFT and hadronization}\label{sec:semeft}

The assumption that the new physics' degrees of freedom lie above the electroweak scale allows to employ the SMEFT in order to parametrize the new-physics effects, which are then encoded in new operators with dimension $D>4$, the first of which arise at $D=6$ if lepton number conservation is assumed. In the following, we assume that the new physics that is relevant to our study appears first via these operators, among which we are only interested on those inducing \CP{} violation in flavor-neutral currents. There are three kinds of such operators: those containing leptons (and photons), those containing quarks and gluons(photons), and those containing quarks and leptons. For the first category, the electron and muon EDMs set strong bounds such that these operators can be ignored~\cite{Panico:2018hal,Sanchez-Puertas:2018tnp}. Concerning the second---hadronic---category, that we name \CP{}$_{\textrm{H}}$, for our process of interest it will manifest as new \CP{}-violating transition form factors, so the $\eta$ coupling to two (generally virtual) photons reads
   \begin{multline}\label{eq:ggvertex}
      i\mathcal{M}^{\mu\nu} = ie^2\Big\{\epsilon^{\mu\nu q_1 q_2}F_{\eta\gamma^*\gamma^*}(q_1^2,q_2^2) + \\
                      \left[ g^{\mu\nu}q_1^2q_2^2 -q_1^2 q_2^{\mu}q_2^{\nu} -q_2^2 q_1^{\mu}q_1^{\nu} 
                                  +  (q_1\!\cdot\!q_2)q_1^{\mu}q_2^{\nu} \right] F_{\eta\gamma^*\gamma^*}^{C\!P2}(q_1^2,q_2^2) \\
                    + \left[g^{\mu\nu}(q_1\!\cdot\!q_2) -q_2^{\mu}q_1^{\nu}\right] F_{\eta\gamma^*\gamma^*}^{C\!P1}(q_1^2,q_2^2)
 \Big\}, 
   \end{multline}
where the connection of the Wilson coefficients to the new form factors $F_{\eta\gamma^*\gamma^*}^{C\!Pi}(q_1^2,q_2^2)$ will be be irrelevant to us after the bounds that the neutron EDM (nEDM) imply are taken into consideration---in the meantime, we shall approximate them as $F_{\eta\gamma^*\gamma^*}^{C\!P1}(q_1^2,q_2^2)\equiv \epsilon_1 F_{\eta\gamma^*\gamma^*}(q_1^2,q_2^2)$ and a similar variant for $F_{\eta\gamma^*\gamma^*}^{C\!P2}(q_1^2,q_2^2)$ (find details in Ref.~\cite{Sanchez-Puertas:2018tnp} and \cite{Escribano:2013kba,Escribano:2015nra,Escribano:2015yup,Masjuan:2015cjl,Masjuan:2017tvw,Sanchez-Puertas:2017sih} for the standard (\CP{}-even $F_{\eta\gamma^*\gamma^*}$) form factor description). 
For the third category, that we name \CP{}$_{\textrm{HL}}$ and mediates quark-lepton interactions, the relevant effective Lagrangian at low energies reads~\cite{Sanchez-Puertas:2018tnp}
\begin{align}
\mathcal{L} ={}& -\mathcal{C}\eta\bar{\mu}\mu, \nonumber \\
\mathcal{C} ={}& \operatorname{Im}[1.57(c_{\ell equ}^{(1)2211}+c_{\ell edq}^{2211}) -2.37  c_{\ell edq}^{2222}]\times 10^{-6}. \label{eq:mathcalC}
\end{align}
where the numeric value is related to the matrix element $\frac{1}{2v^2} \bra{0} \bar{q}^ai\gamma^5 q^a \ket{\eta}$~\cite{Sanchez-Puertas:2018tnp}, connected to the $\eta-\eta'$ mixing parameters~\cite{Feldmann:1999uf,Escribano:2015yup}, and where the involved Wilson coefficients correspond to the $\mathcal{O}_{\ell equ}^{(1)prst}$ and $\mathcal{O}_{\ell edq}^{prst}$ operators---following the same conventions as Refs.~\cite{Grzadkowski:2010es,Falkowski:2017pss}. In the following section, we employ the results in \cref{eq:ggvertex,eq:mathcalC} to estimate their impact in the selected \CP{}-violating $\eta$ muonic decays.

\section{Polarized decays and sensitivities}\label{sec:pol}

A realistic study needs to take into account not only the muon polarization, but the spin analyzing power~\cite{Berge:2011ij} and the asymmetries chosen to probe \CP{} violation.\footnote{As an example, the polarization of the electron cannot be accessed at REDTOP and would be consequently useless.} In this section we compute therefore not only the polarization of the muons, which is all that is required for a subsequent MC implementation with {\textsc{Geant4}}~\cite{Agostinelli:2002hh}, but we estimate in addition the size of the different asymmetries that will be eventually the experimenatlists' target. This size will determine, together with the statistics, the sensitiviy of each process to the Wilson coefficients.

\paragraph{Dimuon decay $\eta\to\mu^+\mu^-$} The dimuon decay emerges as the most competitive decay, as we shall show. The most general amplitude for such decay reads
\begin{equation}
\mathcal{M} = g_P\bar{u} i\gamma^5 v +g_S\bar{u}v.
\end{equation}
In the SM, $g_P = -2m_{\mu}\alpha^2F_{\eta\gamma\gamma}\mathcal{A}(m_{\eta}^2)$ at one-loop via an intermediate two-photon state,\footnote{$F_{\eta\gamma\gamma}$ stands for the normalization of the $\eta$ transition form factor, $F_{\eta\gamma^*\gamma^*}(0,0)=0.2738(47)$~GeV$^{-1}$, while $\mathcal{A} =-1.26 -5.47i$ has been estimated in \cite{Masjuan:2015cjl,Sanchez-Puertas:2017sih} (errors have been ommitted).} while $g_S\simeq 0$. For the \CP{}-conserving contributions, the final leptons are in the $^1S_0$ state, while for the \CP{}-violating one they are in the $^3P_0$, which induces different spin correlations. Indeed, the matrix element for polarized decays reads
  \begin{multline}
    |\mathcal{M}(\SLV{n},\bar{\SLV{n}})|^2 = \frac{m_{\eta}^2}{2} \Big[
      |g_P|^2 \left(1 -  [\SLV{n}\cdot\bar{\SLV{n}}] \right) 
       \\
      +2\left[
         \operatorname{Re}(g_Pg_S^*)(\bar{\SLV{n}}\times \SLV{n})\cdot \SLV{\beta}_{\mu} 
      + \operatorname{Im}(g_Pg_S^*)\SLV{\beta}_{\mu} \cdot(\SLV{n} -\bar{\SLV{n}})
    \right] \\ + |g_S|^2\beta_{\mu}^2 \big(1 -  [ n_z \bar{n}_z  -n_T\cdot\bar{n}_T ]   \big)\Big], 
  \end{multline}
where $\SLV{\beta}_{\mu}$ corresponds to the $\mu^+$ velocity, $\SLV{n}(\bar{\SLV{n}})$ is the spin polarization axis for the $\mu^+(\mu^-)$ in its rest frame, and the $\hat{z}$ axis is defined by the $\mu^+$ direction in the $\eta$ frame; this is the necessary input for MC simulations in  {\textsc{Geant4}}. 
The asymmetries are defined in terms of the $e^+(e^-)$ direction with respect to the $\mu^+$ that, as a consequence of the electroweak interactions, are emmitted preferably along(against) the $\mu^+(\mu^-)$ spin direction, respectively. Defining the standard longitudinal and transverse asymmetries as in Ref.~\cite{Sanchez-Puertas:2018tnp}, we find
   \begin{align}
     A_{L} \equiv&{}  \bar{A}_{L} =
       \frac{\beta_{\mu}}{3} \frac{\operatorname{Im}\mathcal{A} \ \tilde{g}_S}{|\mathcal{A}|^2}, &
   A_{T} \equiv&{}        \frac{\pi\beta_{\mu}}{36} \frac{\operatorname{Re}\mathcal{A} \ \tilde{g}_S}{|\mathcal{A}|^2}, 
   \end{align}
where $\tilde{g}_S = -g_S(2m_{\mu}\alpha^2 F_{\eta\gamma\gamma})^{-1}$. The contribution from our \CP{}$_{\textrm{H,HL}}$-violating scenarios defined in \cref{sec:semeft} yields $g_{S}^{\textrm{HL}} = -\mathcal{C}$, see \cref{eq:mathcalC}, while a loop calculation similar to the SM contribution yields $\tilde{g}_S^{\textrm{H}} = (-0.87 -5.5i)\epsilon_1 + 0.66\epsilon_2$---see Ref.~\cite{Sanchez-Puertas:2018tnp} for details. This results in 
  \begin{align*}
      A_{L}^{H} &{}= 0.11 \epsilon_1 -0.04\epsilon_2,  \quad     A_{T}^{H} = -0.07\epsilon_1 -0.002\epsilon_2, \\  
      A_{L}^{H\!L} &{}= -\operatorname{Im}(2.7(c_{\ell equ}^{(1)2211} +c_{\ell edq}^{2211}) -4.1c_{\ell edq}^{2222})\times 10^{-2}, \\
      A_{T}^{H\!L} &{}= -\operatorname{Im}(1.6(c_{\ell equ}^{(1)2211} +c_{\ell edq}^{2222}) -2.5 c_{\ell edq}^{2222})\times 10^{-3}.
  \end{align*}
With the foreseen statistics at REDTOP and the branching ratio for this decay~\cite{Tanabashi:2018oca}, the statistical noise will be of the order of $3\times 10^{-4}$, that implies sensitivities of the order $\epsilon_{1(2)}\sim 10^{-3(2)}$ and $c_{\mathcal{O}}^{22st}\sim 10^{-2}$.

\paragraph{Dalitz decay $\eta\to\mu^+\mu^-\gamma$} 

For the Dalitz decay, we don't quote the generic matrix element, but the contributions from the SM and \CP{}$_{\textrm{H,HL}}$-violating scenarios---for the particular expressions, we refer the reader to Ref.~\cite{Sanchez-Puertas:2018tnp}. In this case there are additional longitudinal ($A_{L\gamma}$) and transverse ($A_{TL}$) asymmetries connected to the orientation relative to the photon~\cite{Sanchez-Puertas:2018tnp}. For conciseness, we quote here the final results for the asymmetries and refer the reader to Ref.~\cite{Sanchez-Puertas:2018tnp} for details:\footnote{We checked that parity-violating asymmetries connected to an intermediate $Z^*\gamma$ state are irrelevant for the considered asymmetries.}
   \begin{align}
   A_{L}^{H} &{}= A_{T}^{H} = A_{TL}^{H} = 0, \quad A_{L\gamma}^{H} = -0.002\epsilon_1 \\
   A_{L}^{HL} &{}= -4\operatorname{Im}(1.1(c_{\ell equ}^{(1)221} +c_{\ell edq}^{221}) -1.7c_{\ell edq}^{2222})\!\times\! 10^{-7}, \\ 
   A_{T}^{HL} &{}= -5\operatorname{Im}(1.1(c_{\ell equ}^{(1)221} +c_{\ell edq}^{221}) -1.7c_{\ell edq}^{2222})\!\times\! 10^{-6} , \\ 
   A_{L\gamma}^{HL} &{}= 5\operatorname{Im}(1.1(c_{\ell equ}^{(1)2211} +c_{\ell edq}^{2211}) -1.7c_{\ell edq}^{2222})\!\times\! 10^{-6}, \\
   A_{TL}^{HL} &{}= 2\operatorname{Im}(1.1(c_{\ell equ}^{(1)221} +c_{\ell edq}^{221}) -1.7c_{\ell edq}^{2222})\!\times\! 10^{-5}.
   \end{align}
From the branching ratio~\cite{Tanabashi:2018oca} and expected statistics, the noise is expected at the level of $10^{-5}$, achieving sensitiviteis of the order $\epsilon_1\sim 10^{-2}$ and $c_{\mathcal{O}}^{22st}\sim1$---less competitive than dimuon decays.

\paragraph{Double-Dalitz decay $\eta\to\mu^+\mu^-e^+e^-$}

\begin{figure}
  \includegraphics[width=\columnwidth]{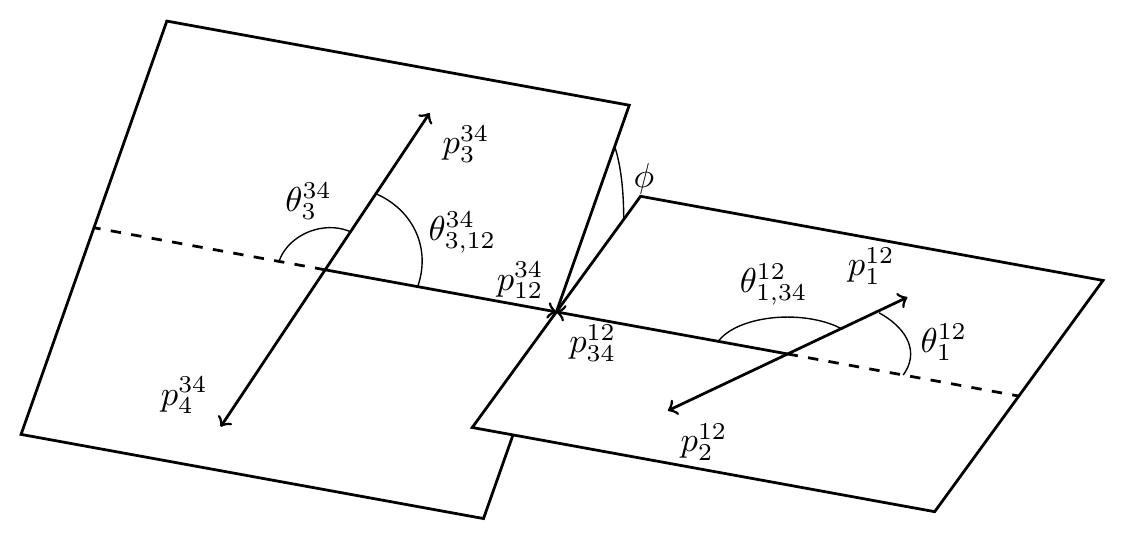}
  \caption{Double-Dalitz decay kinematics. The dilepton systems are shown in their corresponding rest frame. Note that $\phi$ does not change from $\eta$ to dilepton frames.\label{fig:ddalitz}}
\end{figure}

Finally, we consider also the $\eta\to\mu^+\mu^-e^+e^-$ double-Dalitz decay, but we ommit the pure electronic and muonic channels. The reason is that the pure muonic one is extremely suppressed by phase-space, while the electronic case would suffer from yet stronger bounds arising from the electron EDM or parity violation in heavy atoms~\cite{Yanase:2018qqq}. For this decay, polarization observables are not required for probing \CP{}-violating effects; instead, \CP{}-violation is related to the leptons' plane azimuthal angle $\phi\to-\phi$ behavior---see \cref{fig:ddalitz}.
There are different asymmetries that can be defined. In our case, the most interesting one is that related to the fraction of events corresponding to $\sin(2\phi)>0$ as compared to those for which $\sin(2\phi)<0$ ($\phi/2\to-\phi/2$ asymmetry), that we name $A_{\phi\!/2}$ and for which we obtain 
   \begin{align}
      A_{\phi\!/2}^{H} &{}= -0.2\epsilon_1 +0.0003\epsilon_2, \\
      A_{\phi\!/2}^{H\!L} &{}= -\!\operatorname{Im}(1.3(c_{\ell e qu}^{(1)2211} +c_{\ell edq}^{2211}) -1.9c_{\ell edq}^{2222})\!\times\! 10^{-5}. 
   \end{align}
From the branching ratio computed in Ref.~\cite{Kampf:2018wau} and expected statistics, the noise is expected at the level of $5\times 10^{-4}$, achieving sensitiviteis of the order $\epsilon_1\sim 10^{-3}$ and $c_{\mathcal{O}}\sim 40$---with similar performance to dimuon decay for the \CP{}$_{\textrm{H}}$ scenario, but a poor performance for the \CP{}$_{\textrm{HL}}$.

\paragraph{Summary of sensitivities} 

To summarize the findings, we collect the different sensitivities in \cref{tab:bounds}.
\begin{table}[t]
 \resizebox{\linewidth}{!}{
  \begin{tabular}{ccccc} \toprule
    Process & $\epsilon_{1}$ & $\epsilon_{2}$ & Im$\{c_{lequ}^{(1)2211},c_{ledq}^{2211}\} $ & Im$c_{ledq}^{2222}$ \\ \midrule
    $\eta\to\mu^+\mu^-$ & $0.003$ & $0.008$ & $0.01$ & $0.007$ \\
    $\eta\to\gamma\mu^+\mu^-$ & $0.02$ & - & $2$ & $1$ \\
    $\eta\to e^+e^-\mu^+\mu^-$ & $0.003$ & $2$ & $40$& $25$ \\ \bottomrule
  \end{tabular}}
 \caption{The lower bounds (in magnitude) for the sensitivities to the different parameters.\label{tab:bounds}}
\end{table}
Clearly, the most competitive decay is the $\eta\to\mu^+\mu^-$, providing competitive sensitivities for all scenarios, while the Dalitz decay seems the less competitive one in any scenario. The reason is clear: the dimuon decay is suppressed in the SM due to the electromagnetic $(\alpha^2)$ and helicity-flip $(m_{\mu}F_{\eta\gamma\gamma})$ suppresion, while in the \CP{}$_{\textrm{HL}}$ scenario it ocurs at tree level and helicity-flip suppression is absent; by contrast, for the \CP{}$_{\textrm{H}}$ scenario, the suppression is similar to the SM one for all decays.

\section{Bounds and constraints}\label{sec:bounds}

In this section we explore the bounds that \CP{}-violating observables can set on the discussed scenarios. A powerful observable is the electron EDM~\cite{Andreev:2018ayy,Panico:2018hal} and parity-violation in heavy atoms~\cite{Yanase:2018qqq}---one of the reasons that we restricted to the muonic cases only.\footnote{Remind that, for the purely electronic modes, only double-Dalitz decays can be used to probe for \CP{} violation at REDTOP.} Concerning alternative EDM's, the muon one has been measured, though not as precise as the proton and the neutron ones, that are more restrictive then~\cite{Tanabashi:2018oca}. Among the two of them, we find that the neutron sets the most stringent constraints~\cite{Sanchez-Puertas:2018tnp}, to which we restrict in the following.

As an addition to Ref.~\cite{Sanchez-Puertas:2018tnp}, we take advantage that the SMEFT $SU(2)_L$ symmetry allows to connect the flavor-neutral component of our operators to their charged counterpart. Particularly, this allows to connect with $D_s^+$ meson decays in the \CP{}$_{\textrm{HL}}$ scenario, providing bounds similar to the nEDM. 

\subsection{EDM}

In the following, we discuss the bounds arising from the nEDM in each of the two scenarios separately, given its evaluation is qualitatively different for each of them.  

\paragraph{Hadronic \CP{}-violating scenario}

\begin{figure}\centering
   \includegraphics[width=\columnwidth]{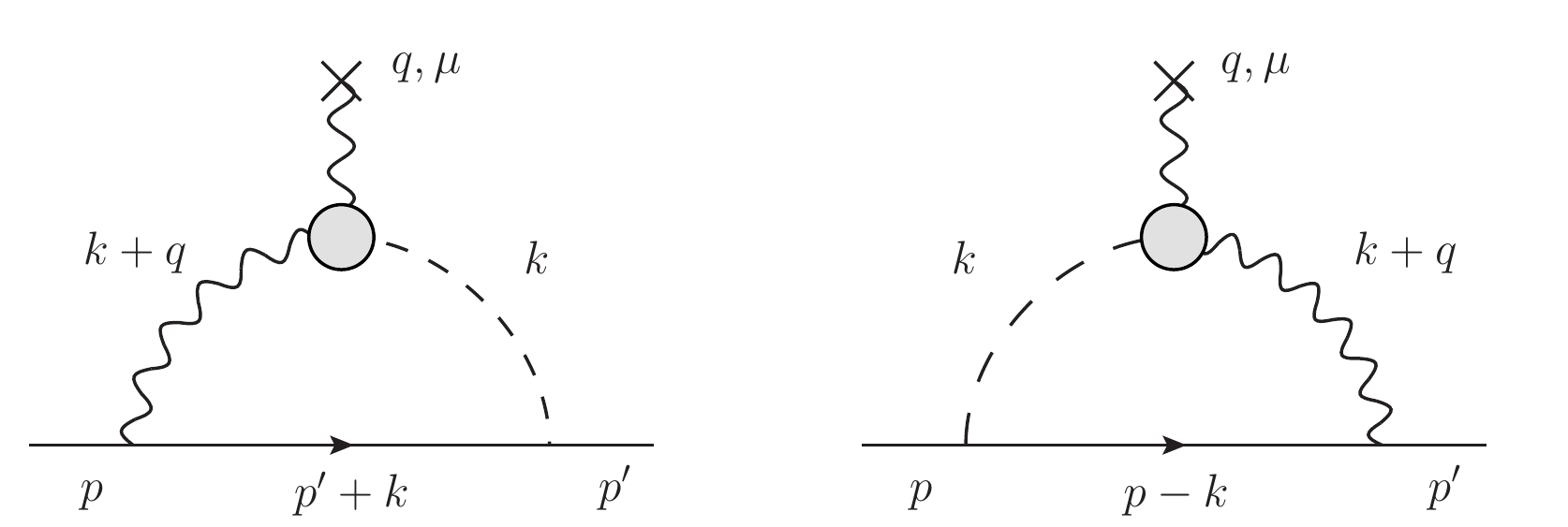}
   \caption{Contributions to the nucleon EDM via a \CP-violating $\eta$ coupling to $\gamma\gamma$. Solid lines represent the nucleon, while dashed lines represent an $\eta$ meson. \label{fig:edm1}}
\end{figure}

In the purely hadronic \CP{}-violating scenario, regardless of the particular underlying microscopic origin (e.g., the particular SMEFT operator selected), there will always exist a contribution such as the one that is depicted in \cref{fig:edm1}.\footnote{We emphasize that, in this scenario, there will be \textit{a priori} more relevant contributions not involving electromagnetic interactions. However, this one will serve to all scenarios (regardless their microscopic origin) and turns out to be restrictive enough to discard them all for the sensitivities accessible at REDTOP, while avoiding a case-by-case analysis.} Employing the $\eta$-coupling to the neutron from $\chi$PT and assuming on-shell electromagnetic form factors (along the same lines as Ref.~\cite{Gutsche:2016jap}), we find that $d_E^n = -6.2\times 10^{-20} \epsilon_1 e~\textrm{cm}$.\footnote{Since the external photon needs to be on-shell, the nEDM places no restriction on $\epsilon_2$.} The current bound for the nEDM ($ d_E^n < 3\times 10^{-26} e~\textrm{cm}$) implies then values for $\epsilon_1$ that are out of reach for REDTOP. Concerning the $\epsilon_2$ parameter, we expect similar bounds since we find no dynamical principle that would prevent an (almost) vanishing coupling to real photons only. As such, we do not expect to observe a \CP{}-violating $F_{\eta\gamma^*\gamma^*}$ transition form factor that has been discussed in the past---especially in the context of the $\pi^0$, that can be ruled out as well.

\paragraph{Hadron-lepton \CP{}-violating scenario}

\begin{figure*}[t]\begin{center}
\includegraphics[width=0.8\textwidth]{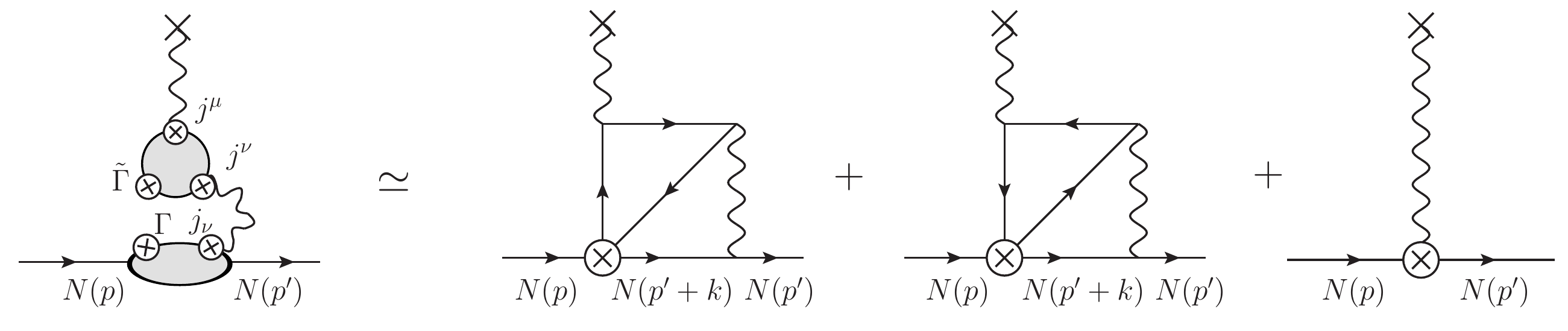}
\caption{The nEDM contribution for the second \CP{}$^{HL}$ scenario (left). The gray blob stands for the required hadronic form factor that is defined in the text. This is approximated via a low-energy part saturated by intermediate neutron states (two additional diagrams are implicit) and a high-energy part, saturated via the OPE and requiring renormalization.\label{fig:nEDM}}
\end{center}\end{figure*}

In the second scenario, the nEDM appears first at the two-loop level, as it is shown in \cref{fig:nEDM}. The resulting contribution to the  $\bra{N(p')} j^{\mu} \ket{N(p)} \equiv  \bar{u}_{p'}\Gamma^{\mu}u_p $ matrix element, where $j^{\mu}$ represents the electromagnetic current, can be expressed as
 \begin{multline}
   \Gamma^{\mu} = \sum_i \! \int \frac{d^4 k}{(2\pi)^4} \frac{e^2}{k^2} 
                   \frac{1}{i}  \!\int \!\! e^{i k\cdot z} \bra{N_{p'}}
                       T\{  j_{\nu}(z) \bar{q}\Gamma_i q(0) \} \ket{N_p}  \\
                    \times \frac{1}{i}\int \! e^{-i (q\cdot x +k\cdot y)} 
                       \bra{0}T\{ j^{\mu}(x)j^{\nu}(y)\bar{\ell}\tilde{\Gamma}_i \ell(0) \} \ket{0},   
 \end{multline}
where $\sum_i\{\Gamma_i,\tilde{\Gamma}_i\} = -\frac{c_{\ell equ}^{(1)}(c_{\ell edq})}{2v^2} [ \{ 1, i\gamma^5 \} \pm \{ i\gamma^5, 1  \} ] $.
In order to extract the contribution to the nEDM, we do an approximation assuming this can be splitted in two different contributions: the first one---at low energies---and dominated by an intermediate neutron state; the second one---at high energies---and approximated via the operator product expansion (OPE); see also \cref{fig:nEDM}. 
For the first one, we approximate at low energies the neutron pseudoscalar $\bra{n} \bar{q}i\gamma^5 q \ket{n}$ form factor via intermediate pseudo-Goldstone boson exchanges $(\pi^0,\eta,\eta')$, while the scalar form factor $\bra{n} \bar{q}q \ket{n}$ is saturated via the lowest-lying scalar resonances, in the lines of Ref.~\cite{Arriola:2012vk}, and normalized to the $\sigma$-terms~\cite{Varnhorst:2015hfk,Cirigliano:2017tqn,Hoferichter:2015dsa,Alarcon:2011zs,Fernando:2018jrz} (find details in Ref.~\cite{Sanchez-Puertas:2018tnp}). 
For the second one, one has to renormalize the effective operators, that involves in addition the $\mathcal{O}_{\ell equ}^{(3)prst}$ and $\mathcal{O}_{u\gamma, d\gamma}$ operators. The resulting contribution is obtained then via the renormalization-group equations (in good agreement to the recent results in Ref.~\cite{Panico:2018hal}), that are taken from the electroweak scale down to some hadronic scale, for which $\mu=2$~GeV is chosen.\footnote{We note that the first (low-energy) part saturates below such scale.} The matrix element that is then required, $\bra{n} \bar{q}\sigma^{\mu\nu}\gamma^5 q \ket{n}$,  is taken from the lattice results in Ref.~\cite{Bhattacharya:2015esa}. Numerically we find 
  \begin{multline}
    d_E^n = \operatorname{Im}
       (-0.75c_{\ell equ}^{(1)2211} +0.92c_{\ell edq}^{2211} +0.08c_{\ell edq}^{2222} 
\\  -0.59c_{\ell equ}^{(1)2211} +0.15c_{\ell edq}^{2211} +0.001c_{\ell edq}^{2222})\times10^{-23}
,\label{eq:led}
  \end{multline}
where the first (second) line corresponds to the low(high)-energy contribution. Again, the results are to be taken as an order of magnitude estimate. Together with the current bound, our results imply that
  \begin{equation}
    | \operatorname{Im}c_{\ell equ}^{(1)2211} | < 0.002, 
    | \operatorname{Im}c_{\ell edq}^{2211} | < 0.003, 
    | \operatorname{Im}c_{\ell edq}^{2222} | < 0.04. 
  \end{equation}
Comparing to \cref{tab:bounds}, we find that current nEDM bounds allow for \CP{}-violating effects only for the $\mathcal{O}_{\ell edq}^{2222}$ operator, that could be accessed in $\eta\to\mu^+\mu^-$ decays. Stated differently,  $\eta\to\mu^+\mu^-$ decays at REDTOP would be complementary to those of nEDM measurements, especially since they are sensitive to a particular flavor configuration.

\subsection{$D_s^-\to\mu\bar{\nu}_{\mu}$ decays}\label{sec:DsDec}

For the surviving \CP{}$_{\textrm{HL}}$ scenario, the $SU(2)_L$ symmetry connects neutral and charged current processes, allowing to link our study to weak decays, as it is the case of $D_s^-\to\mu\bar{\nu}_{\mu}$. Adapting the results for the branching ratio (BR) in Ref.~\cite{Alonso:2016oyd} to our case,\footnote{This requires to replace $B^-\to D_s^-$, $V_{cb}\to V_{cs}$ and the Wilson coefficients, $\epsilon_P\to (c_{\ell equ}^{(1)2222*} -c_{\ell edq}^{2222*})/(2|V_{cs}|)$ in their expressions.} we find that  
\begin{multline}
\textrm{BR}(D_s^-\to\ell\bar{\nu}_{\ell}) = \tau_{D_s^-} \frac{m_{D_s} m_{\ell}^2 f_{D_s}^2 G_F^2 |V_{cs}|^2}{8\pi} \\ \times \left\vert 1 + \frac{m_{D_s}^2( c_{\ell equ}^{(1)2222*} -c_{\ell edq}^{2222*} )}{2|V_{cs}| m_{\ell}(m_c +m_s)}\right\vert^2 .
\end{multline}
Even if the SM interference with the new \CP{}-violating part vanishes for the BR, and \CP{}-violation could not be tested explicitly, the latter would still contribute quadratically to the BR.\footnote{Formally, $D=8$ operators should be accounted for as well for consistency. However, it is sensible that such a quadratic contribution alone should not exceed the current bounds. The author acknowledges Javier Fuentes-Mart{\'i}n and participants to the III-CAFPE Christmas Workshop 2018 at Granada for calling me the attention to this point.} Assuming that the real part is negligible, and taking that $\textrm{BR}(D_s^-\to\ell\bar{\nu}_{\ell}) = 5.50(23)\times 10^{-3}$~\cite{Tanabashi:2018oca}, we find that $\left\vert \operatorname{Im}c_{\ell edq}^{2222} \right\vert < 0.02 $. The bound is similar to the nEDM and the previous conclusion remains the same.

\section{Conclusions and Outlook}

In this work we have studied the possibility to test \CP{}-violation in $\eta$ decays containing muons at REDTOP. This is possible in our study, mainly, thanks to the ability of the proposed experiment to measure the polarization of the muons. Assuming that \CP{}-violating effects beyond the SM appear via new heavy degrees of freedom, the SMEFT is employed, that allows a convenient connection to different observables. In particular, it allows to connect with the nEDM and $D_s^-\to\mu\bar{\nu}_{\mu}$ decays. As a result, we find that it is possible to find \CP{}-violation in $\eta\to\mu^+\mu^-$ decays with the statistics foreseen at REDTOP while avoiding such constraints---with a single SMEFT operator as the plausible source of it. With the current results at hand, it might be interesting in the future to study whether such operator could induce \CP{}-violating effects in $\eta\to\pi^0\mu^+\mu^-$ decays large enough to be observed at REDTOP. Also, it might be interesting to further investigate loop effects that could mix muonic and electronic operators via the renormalization group since tighter bounds exist for the latter.

\section*{Acknowledgements}

The author acknowledges the support received from the Ministerio de Ciencia, Innovaci\'on y Universidades under the grant SEV-2016-0588, the grant
754510 (EU, H2020-MSCA-COFUND-2016), and the grant FPA2017-86989-P, as well as by Secretaria d'Universitats i Recerca del Departament d'Economia i Coneixement de la Generalitat de Catalunya under the grant 2017 SGR 1069. This work was also supported by the Czech Science Foundation (grant no. GACR 18-17224S) and by the project UNCE/SCI/013 of Charles University.

\section*{References}

\bibliography{references}

\end{document}